\documentclass[useAMS,usenatbib]{mn2e}
\usepackage{times}
\usepackage{graphicx}
\usepackage{amssymb}

\title[The dwarf galaxy population in the Perseus Cluster]{Keck spectroscopy of the faint dwarf elliptical galaxy population in the Perseus Cluster core: mixed stellar populations and a flat luminosity function}
\author[S. J. Penny and C. J. Conselice]{Samantha J. Penny$^1$ and Christopher J. Conselice$^1$ \footnotemark[0]\\ 
$^1$School of Physics \& Astronomy, University of Nottingham, Nottingham, NG7 2RD, United Kingdom}

\begin{document}

\maketitle

\begin{abstract}
We present the result of a photometric and Keck-LRIS spectroscopic study of dwarf galaxies in the core of the Perseus Cluster, down to a magnitude of M$_{\rm B}$ $= -12.5$. Spectra were obtained for twenty-three dwarf-galaxy candidates, from which we measure radial velocities and stellar population characteristics from absorption line indices. From radial velocities obtained using these spectra we confirm twelve systems as cluster members, with the remaining eleven as non-members. Using these newly confirmed cluster members, we are able to extend the confirmed colour-magnitude relation for the Perseus Cluster down to M$_{\rm B}$ $= -12.5$. We confirm an increase in the scatter about the colour magnitude relationship below M$_{\rm B}$ $= -15.5$, but reject the hypothesis that very red dwarfs are cluster members. We measure the faint-end slope of the luminosity function between M$_{\rm B}$ $= -18$ and M$_{\rm B}$ $= -12.5$, finding $\alpha$ $= -1.26$ $\pm$  0.06, which is similar to that of the field. This implies that an overabundance of dwarf galaxies does not exist in the core of the Perseus Cluster. By comparing metal and Balmer absorption line indices with $\alpha$-enhanced single stellar population models, we derive ages and metallicities for these newly confirmed cluster members. We find two distinct dwarf elliptical populations: an old, metal poor population with ages $\sim$ 8 Gyr and metallicities $[{\rm Fe/H}]$ $<$ $-0.33$, and a young, metal rich population with ages $<$ 5 Gyr  and metallicities $[{\rm Fe/H}]$ $>$ $-0.33$. Dwarf galaxies in the Perseus Cluster are therefore not a simple homogeneous population, but rather exhibit a range in age and metallicity.
\end{abstract}

\begin{keywords}
galaxies: dwarf -- galaxies: clusters: general -- galaxies: clusters: individual: Perseus Cluster -- galaxies: luminosity function, mass function
\end{keywords}

\section{Introduction}

Faint, low mass galaxies are the most numerous galaxy type in the universe, and are thus fundamental in understanding galaxy formation. However, their low luminosities and low surface brightness make detailed studies of them difficult. Because dwarfs are so common, any galaxy evolution/formation theory must clearly be able to predict and describe the properties of these galaxies.

The luminosity functions of nearby galaxies in all environments reveal that dwarf galaxies are far more common than brighter galaxies.  Within galaxy clusters there furthermore appears to be an overdensity of dwarf galaxies when compared to the field. The origin of these extra dwarfs, or if this excess is even real, remains unknown. The cluster luminosity function itself is not universal, but is strongly dependent on environment \citep{sab03}, with the luminosity function often steeper in the more diffuse outer regions of clusters than in the denser cluster cores. The luminosity function also depends on the individual cluster, with each one having a characteristic luminosity function. For example, the Fornax Cluster is compact and rich in early-type galaxies, with a flat luminosity function faint-end slope of $\alpha = -1.1$ \citep{mie}. The Virgo Cluster has a high abundance of spiral galaxies and has a steeper luminosity function with a faint end slope  of $\alpha = -1.6$ (\citealt{trenth02,sab03}). 

We know that both Local Group (LG) dwarf elliptical and dwarf spheroidal galaxies display varying star formation histories, with metal poor populations as old as the classical halo globular galaxies, but with evidence for star formation as little as 2-3 Gyr ago. Some low-mass LG galaxies such as Sagittarius, contain surprisingly high metal rich stellar populations considering their luminosities, as is also seen in clusters of galaxies. Previous spectroscopic and ground based imaging has revealed that dwarfs in the core of clusters are not a simple homogeneous population, with cluster core galaxies fainter than M$_{\rm B}$ $= -15$ containing a mix of metallicities and ages (\citealt*{Po,RSMP,c1}). 

The formation scenarios for dwarf galaxies fall into two categories. The first of these is that dwarfs are old, primordial objects (hierarchical model) (e.g. \citealt{wf}). The second is that they have recently evolved or transformed from a progenitor galaxy population (e.g. \citealt*{moore}). Within hierarchical models of galaxy formation, dwarf galaxies are formed from small density fluctuations in the early universe. The lowest mass systems (i.e. dwarf galaxies) form first, and massive galaxies are built from these in mergers. If star formation occurred soon after the gravitational collapse of initial density perturbations, dwarf galaxy halos would be amongst the first objects to form (e.g. \citealt{ds}). Theory also predicts that such systems would form first in the densest environments, although cluster dEs would not necessarily contain the first stars formed in the universe. These halos could have formed their stars quite recently, if star formation was suppressed in some way \citep{tully}. 

Dwarfs in clusters, on the other hand, could also be the remnants of stripped discs or dwarf irregulars. Through galaxy harassment and interactions, these progenitors become morphologically transformed into dEs. Cluster galaxies become stripped of their interstellar medium, and become dynamically heated by high-speed interactions with other galaxies and the gravitational potential of the cluster. To compensate, the galaxy loses stars,  and over time the spiral can morphologically transform into a dE \citep*{c3}. This model is supported by recent observations of embedded discs in dEs in the Virgo and Fornax clusters (\citealt*{Bar,DeR}). Distant clusters at z $\approx$ 0.4 are also filled with small spiral galaxies, but this population is largely absent in nearby clusters, where dwarf spheroidals make up the faint-end of the luminosity function \citep{moore}.
 
Downsizing \citep{DeL} is the observed trend that star formation occurs later, and over more extended timescales, in smaller galaxies. In this scenario, dEs and lower mass galaxies formed or entered clusters after the giant galaxies. Low mass galaxies on average end their star formation after the giant elliptical galaxies. For example, the faint end of the red sequence in clusters is not formed until z $<$ 0.8 \citep{DeL}, implying the luminosity weighted stellar populations of lower mass galaxies are younger than the stars in the giant elliptical galaxies. This seems to contradict the hierarchical method of galaxy formation, where dwarf systems form first. Also, dwarf elliptical galaxies are preferentially found in dense cluster environments, with few examples of isolated dEs, again contradicting the hierarchical model of galaxy formation. However, some dEs in clusters contain old stellar populations, so it is likely that multiple formation methods are needed to explain the origin of cluster dwarfs.

The colour-magnitude relation (CMR) for cluster galaxies is well defined at bright magnitudes and forms a tight sequence. However, it is unclear what the shape of the red sequence is at fainter magnitudes \citep{an}. \citet*{c2} find that galaxies at the brightest 4 magnitudes of Perseus obey the CMR, with the fainter candidate members showing significant scatter from the relation. In contrast, \citet{an} and others find no deviation from the CMR at fainter magnitudes in other clusters. Instead, the faint red sequence is an extrapolation of what is observed at brighter magnitudes. With spectroscopy, we can test this directly by finding confirmed dwarf members of the cluster to establish the true form of the colour magnitude relation.

Spectroscopy can also be used to measure the faint-end of the luminosity function in galaxy clusters, which is fundamental in describing the galaxy population. The luminosity function contains important information on the formation and evolution of galaxies, and at low redshifts contains the combined influence of the galaxy initial mass function, and the effects of any evolutionary processes that have taken place in the cluster since its formation.

In this paper, we use deep Keck spectroscopy to determine cluster membership for dwarfs at M$_{\rm B}$ $<$ $-16$, and present the ages and metallicities of these galaxies based on the strengths of  Balmer and metal absorption lines in their spectra. We confirm that twelve dwarfs are Perseus Cluster members, while the remaining objects are background galaxies. Using these results, we examine the colour-magnitude relation for the Perseus Cluster down to M$_{\rm B}$ $= -12.5$. We also measure the luminosity function faint-end slope $\alpha$ based on our cluster membership results. Our main conclusion is that some dwarf galaxies in Perseus have old, metal poor populations, whilst others are younger, metal rich systems. This suggests that dwarf galaxies in the core of the Perseus Cluster are not a simple, homogeneous population, but require multiple scenarios for their formation.

This paper is organised as follows: in $\S$ 2 we discuss the observations and the selection criteria for the dwarf galaxies, $\S$ 3.1 identifies the cluster members, $\S$ 3.2 presents the colour magnitude relation, and the central luminosity function is presented in $\S$ 3.3. In $\S$ 3.4 we derive luminosity weighted ages and metallicities for the newly confirmed cluster members. A discussion of the main results is presented in $\S$ 4 and these results are summarised in $\S$ 5. We assume the distance to the Perseus Cluster is 77 Mpc throughout this paper.

\section{Observations}

The Perseus Cluster (Abell 426) is one of the richest nearby galaxy clusters, with a redshift $v$ = 5366 km s$^{-1}$ \citep{sr}, and at a distance D = 77 Mpc. It also has a high velocity dispersion of $\sigma$ = 1324 km s$^{-1}$ \citep{sr}. Despite its proximity it has not been studied in as much detail as clusters such as Virgo and Fornax, most likely due to its low Galactic latitude (\textit{b} = $-13^{\circ}$). 

Candidate dwarf elliptical galaxies in Perseus for our Keck spectroscopic observations were selected from those listed in \citet{c1}, and we follow the same numbering system throughout this paper. All dwarfs in this list have colours (B$-$R)$_{0}$ $<$ 2, and are symmetric, round or elliptical in shape, without evidence for star formation. All the candidates also have a central surface brightness fainter than $\mu_{\rm B}$ = 24.0 mag arcsec$^{-2}$, and have near-exponential surface brightness profiles. We restrict our study to those dwarf galaxies with M$_{\rm B}$ $<$ $-12.5$ to be more certain of cluster membership. For more detail of the selection process, see \citet{c1} and \citet{c2}. 

We also observed an additional two objects not in this original list, one of which is a dwarf galaxy, and the other an irregular galaxy (selected to fill the slits on one of the masks during spectroscopy). We also carried out photometry on these additional objects so they could be included on the colour-magnitude diagram presented in $\S$ 3.2. All candidates lie within the central 173 arcmin$^2$ of the cluster, and have M$_{\rm B}$ $= -16$ to M$_{\rm B}$ $= -12.5$. 

\begin{table*}
\begin{minipage}{150mm}
\caption{Candidate dwarf galaxies in the Perseus Cluster core}
\begin{tabular}{lcccccc}
\hline
 & $\alpha$  & $\delta$ & B & (B$-$R) & $v{\rm_{radial}}$ & \\
 Galaxy & (J2000.0) & (J2000.0) & (mag) & (mag) & (km s$^{-1}$) & Cluster Member \\
\hline
CGW 1*  & 03 18 53.5 & +41 31 50.1 & 20.54 $\pm$ 0.02 & 0.91 $\pm$ 0.05 & \ldots & \ldots \\
CGW 2*  & 03 18 53.8 & +41 32 52.3 & 20.01 $\pm$ 0.01 & 1.14 $\pm$ 0.05 & \ldots & \ldots \\
CGW 3   & 03 18 54.9 & +41 25 55.3 & 21.61 $\pm$ 0.04 & 1.80 $\pm$ 0.05 & \ldots & \ldots \\
CGW 4*  & 03 18 55.7 & +41 25 54.4 & 20.76 $\pm$ 0.03 & 1.44 $\pm$ 0.04 & \ldots & \ldots \\
CGW 5   & 03 18 55.8 & +41 23 41.3 & 21.46 $\pm$ 0.03 & 1.55 $\pm$ 0.05 & 80585 $\pm$ 182 & N \\
CGW 6   & 03 18 57.8 & +41 24 57.6 & 20.06 $\pm$ 0.01 & 1.22 $\pm$ 0.04 & \ldots & \ldots \\
CGW 7   & 03 18 59.5 & +41 31 18.9 & 21.39 $\pm$ 0.04 & 0.89 $\pm$ 0.07 & \ldots & \ldots \\
CGW 8   & 03 18 59.7 & +41 26 40.0 & 19.71 $\pm$ 0.01 & 1.32 $\pm$ 0.04 & \ldots & \ldots \\
CGW 9   & 03 18 59.8 & +41 30 37.1 & 21.34 $\pm$ 0.02 & 0.75 $\pm$ 0.05 & \ldots & \ldots \\
CGW 10  & 03 19 00.4 & +41 29 02.4 & 20.26 $\pm$ 0.02 & 1.29 $\pm$ 0.05 & 5845 $\pm$ 47 & Y \\
CGW 11* & 03 19 03.5 & +41 35 13.6 & 20.21 $\pm$ 0.02 & 1.09 $\pm$ 0.05 & \ldots & \ldots \\
CGW 12* & 03 19 04.2 & +41 35 20.6 & 20.95 $\pm$ 0.02 & 1.15 $\pm$ 0.04 & \ldots & \ldots \\
CGW 13* & 03 19 04.7 & +41 32 24.6 & 21.50 $\pm$ 0.02 & 1.95 $\pm$ 0.05 & \ldots & \ldots \\
CGW 14  & 03 19 05.2 & +41 25 07.0 & 21.87 $\pm$ 0.04 & 1.58 $\pm$ 0.06 & \ldots & \ldots \\
CGW 15* & 03 19 05.2 & +41 34 48.1 & 20.37 $\pm$ 0.02 & 1.09 $\pm$ 0.05 & \ldots & \ldots \\
CGW 16  & 03 19 06.0 & +41 26 18.7 & 19.78 $\pm$ 0.02 & 1.26 $\pm$ 0.05 & 6577 $\pm$ 45 & Y \\
CGW 17  & 03 19 06.8 & +41 26 40.8 & 20.86 $\pm$ 0.05 & 1.70 $\pm$ 0.06 & \ldots & \ldots \\
CGW 18  & 03 19 09.1 & +41 32 41.7 & 21.57 $\pm$ 0.02 & 1.01 $\pm$ 0.04 & 46906 $\pm$ 210 & N \\
CGW 19  & 03 19 09.5 & +41 31 30.9 & 20.59 $\pm$ 0.01 & 0.91 $\pm$ 0.04 & 67527 $\pm$ 114 & N \\
CGW 20  & 03 19 10.4 & +41 29 37.0 & 19.27 $\pm$ 0.01 & 1.27 $\pm$ 0.04 & 7325 $\pm$ 48 & Y\\
CGW 21  & 03 19 12.7 & +41 30 37.7 & 21.61 $\pm$ 0.04 & 1.00 $\pm$ 0.07 & 8741 $\pm$ 31 & Y\\
CGW 22* & 03 19 13.0 & +41 34 51.8 & 20.10 $\pm$ 0.01 & 1.72 $\pm$ 0.04 & \ldots & \ldots \\
CGW 23* & 03 19 14.9 & +41 30 27.1 & 21.77 $\pm$ 0.06 & 0.85 $\pm$ 0.11 & \ldots & \ldots \\
CGW 24  & 03 19 15.1 & +41 28 56.6 & 21.40 $\pm$ 0.05 & 1.12 $\pm$ 0.04 & 64780 $\pm$ 146 & N\\
CGW 25  & 03 19 15.9 & +41 30 20.3 & 21.89 $\pm$ 0.05 & 1.11 $\pm$ 0.08 & 70320 $\pm$ 59 & N \\
CGW 26* & 03 19 17.3 & +41 34 54.5 & 21.97 $\pm$ 0.04 & 1.63 $\pm$ 0.06 & \ldots & \ldots \\
CGW 27  & 03 19 17.6 & +41 29 57.7 & 21.40 $\pm$ 0.05 & 1.64 $\pm$ 0.05 & 125580 $\pm$ 51 & N\\
CGW 28  & 03 19 18.0 & +41 24 25.3 & 21.62 $\pm$ 0.03 & 1.25 $\pm$ 0.05 & \ldots & \ldots \\
CGW 29  & 03 19 18.8 & +41 26 32.4 & 20.94 $\pm$ 0.03 & 1.02 $\pm$ 0.06 & 5110 $\pm$ 77 & Y \\
CGW 30  & 03 19 20.1 & +41 24 06.6 & 21.34 $\pm$ 0.04 & 1.57 $\pm$ 0.06 & \ldots & \ldots \\
CGW 31  & 03 19 21.2 & +41 28 42.5 & 20.77 $\pm$ 0.03 & 1.40 $\pm$ 0.06 & 8662 $\pm$ 14 & Y\\
CGW 32* & 03 19 22.1 & +41 24 27.2 & 20.15 $\pm$ 0.01 & 1.16 $\pm$ 0.04 & \ldots & \ldots \\
CGW 33  & 03 19 22.4 & +41 24 04.2 & 21.75 $\pm$ 0.05 & 1.90 $\pm$ 0.07 & \ldots & \ldots \\
CGW 34* & 03 19 24.6 & +41 25 53.4 & 21.75 $\pm$ 0.05 & 1.29 $\pm$ 0.06 & \ldots & \ldots \\
CGW 35  & 03 19 24.7 & +41 24 36.3 & 21.90 $\pm$ 0.07 & 1.65 $\pm$ 0.08 & \ldots & \ldots \\
CGW 36* & 03 19 24.7 & +41 25 52.6 & 21.80 $\pm$ 0.03 & 1.48 $\pm$ 0.05 & \ldots & \ldots \\
CGW 37  & 03 19 26.9 & +41 33 05.1 & 20.74 $\pm$ 0.03 & 1.43 $\pm$ 0.06 & 70960 $\pm$ 23 & N \\
CGW 38  & 03 19 27.1 & +41 27 16.1 & 18.48 $\pm$ 0.01 & 1.43 $\pm$ 0.04 & 4151 $\pm$ 59 & Y\\ 
CGW 39  & 03 19 31.4 & +41 26 28.7 & 18.72 $\pm$ 0.01 & 1.39 $\pm$ 0.04 & 6421 $\pm$ 50 & Y \\
CGW 40  & 03 19 31.7 & +41 31 21.3 & 19.30 $\pm$ 0.01 & 1.24 $\pm$ 0.04 & \ldots & \ldots \\
CGW 41  & 03 19 33.5 & +41 24 39.4 & 21.76 $\pm$ 0.08 & 1.55 $\pm$ 0.10 & \ldots & \ldots \\
CGW 42  & 03 19 36.1 & +41 32 47.4 & 21.82 $\pm$ 0.03 & 1.01 $\pm$ 0.05 & 51225 $\pm$ 148 & N \\
CGW 43* & 03 19 38.6 & +41 33 46.1 & 21.92 $\pm$ 0.03 & 1.03 $\pm$ 0.05 & \ldots & \ldots \\
CGW 44  & 03 19 39.7 & +41 26 39.1 & 21.05 $\pm$ 0.02 & 1.21 $\pm$ 0.05 & \ldots & \ldots \\
CGW 45  & 03 19 41.7 & +41 29 17.0 & 18.72 $\pm$ 0.01 & 1.42 $\pm$ 0.04 & 3120 $\pm$ 27 & Y \\
CGW 46  & 03 19 42.3 & +41 34 16.6 & 20.67 $\pm$ 0.02 & 1.04 $\pm$ 0.05 & \ldots & \ldots \\
CGW 47  & 03 19 43.5 & +41 28 52.9 & 20.87 $\pm$ 0.02 & 1.24 $\pm$ 0.05 & 2886 $\pm$ 48 & Y \\
CGW 48  & 03 19 45.7 & +41 32 18.0 & 21.56 $\pm$ 0.05 & 0.44 $\pm$ 0.11 & \ldots & \ldots \\
CGW 49  & 03 19 46.1 & +41 24 51.2 & 21.43 $\pm$ 0.03 & 1.41 $\pm$ 0.05 & 8457 $\pm$ 38 & Y \\
CGW 50* & 03 19 55.4 & +41 25 04.4 & 21.06 $\pm$ 0.02 & 1.87 $\pm$ 0.05 & \ldots & \ldots \\
CGW 51  & 03 19 56.1 & +41 29 09.4 & 20.71 $\pm$ 0.01 & 1.42 $\pm$ 0.04 & 69132 $\pm$ 82 & N \\
CGW 52  & 03 19 56.1 & +41 32 38.7 & 21.31 $\pm$ 0.02 & 1.87 $\pm$ 0.05 & 40994 $\pm$ 105 & N \\
CGW 53  & 03 19 58.5 & +41 31 02.0 & 21.86 $\pm$ 0.03 & 1.00 $\pm$ 0.06 & 64495 $\pm$ 79 & N \\
54      & 03 19 48.6 & +41 33 28.6 & 19.75 $\pm$ 0.03 & 1.30 $\pm$ 0.04 & 3363 $\pm$ 38 & Y \\
Irregular & 03 19 05.1 & +41 28 12.4 & 15.19 $\pm$ 0.01 & 1.41 $\pm$ 0.05 & 4027 $\pm$ 39 & Y \\
\hline
\end{tabular}

\medskip
* Not targeted for spectroscopy\\
Units of right ascension are hours, minutes and seconds, and units of declination are degrees, arcminutes and arcseconds. Magnitudes and colours are taken from \citet{c1}. 54 and the irregular galaxies are additional objects to the original list in \citet{c1}. Galaxies flagged with an * were not targeted in our spectroscopy, but are included in the colour-magnitude relation and luminosity function for the Perseus Cluster.
\end{minipage}
\end{table*} 

The imaging data from which candidate dwarfs are acquired was taken using the WIYN 3.5m f/6.2 telescope located at the Kitt Peak National Observatory. Harris $B$ and $R$ images were acquired on the nights of 1998 November 14 and 15 under photometric conditions, with an average seeing of 0.7''. Exposure times were 800s in the $R$ band and 1000s in $B$. Flat fields were taken prior to each night of observing, and combined to create a single flat. Landolt standard star fields were obtained  throughout each night, allowing us to find accurate zero points, air masses and colour terms. See \citet{c2} and \citet{c1} for details of the data reduction. 

The new spectroscopic observations we present in this paper were taken using the Keck Low Resolution Imaging Spectrometer (LRIS) \citep{oke} on the Keck I Telescope on 2002 November 26 and 27. Using four LRIS masks, we targeted forty-one dwarf galaxies, and one irregular, and potentially interacting, galaxy in the Perseus Cluster. 

Successful spectra were obtained for twenty-three dwarf galaxy candidates from \citet{c1}, and the potentially interacting galaxy. The spectra were taken with an average seeing of 1.1'' under clear, but non-photometric, conditions. We obtained spectroscopy with both the red and blue sides of LRIS. On the blue side we used a 400 line mm$^{-1}$ grism blazed at 3400 {\rm \AA}, whilst on the red side we used a 600 line mm$^{-1}$ grating blazed at 5000 {\rm \AA}, producing  effective resolutions of 8 and 6 {\rm \AA}, respectively. The spectra were bias subtracted, flat-fielded and rectified using IRAF\footnote{IRAF is distributed by the National Optical Astronomy Observatories which are operated by the Association of Universities for Research in Astronomy, Inc., under cooperative agreement with the National Science Foundation} reduction techniques for multi-slit data. This combined the various frames into a single one, from which the spectra for the individual dwarf galaxies could be extracted separately. 

The positions and photometric properties of all our science objects are listed in Table 1. The radial velocities in Table 1 are generally measured on the red spectra, where the wavelength calibration was more reliable, and include heliocentric corrections. The errors on the radial velocities are based on uncertainties in the wavelength calibration, and determined by measuring the wavelengths of skylines in the spectra and comparing these to expected values. The difference in the actual and expected positions of the skylines were then converted into a velocity error.

We also obtained \textit{Hubble Space Telescope} ($HST$) Advanced Camera for Surveys (ACS) imaging in the F555W and F814W bands (Penny et al. 2008, in preparation), from which we obtain high resolution images of several of our cluster dwarfs and non-members. 

\section{Results}

\subsection{Cluster membership}

Using our spectra we identify absorption features such as the Mg $b$ and Fe5270 lines, as well as Balmer lines such as H$\beta$. From these lines we derive the radial velocities of each of the dwarfs through comparing the measured absorption lines with those of known early-type galaxy spectra by identifying well known lines by eye.

\begin{table}
\caption{Potential background cluster members}
\centering
\begin{tabular}{@{}lccc}
\hline
 & $\alpha$  & $\delta$  & \\
Galaxy & (J2000.0) & (J2000.0) & $z$ \\
\hline
CGW 19 & 03 19 09.5 & +41 31 30.9 & 0.22\\
CGW 24 & 03 19 15.1 & +41 28 56.6 & 0.23\\
CGW 25 & 03 19 15.9 & +41 30 20.3 & 0.23\\
CGW 37 & 03 19 26.9 & +41 33 05.1 & 0.24\\
CGW 51 & 03 19 56.1 & +41 29 09.4 & 0.23\\
CGW 53 & 03 19 58.5 & +41 31 02.0 & 0.22\\
\hline
\end{tabular}
\end{table} 

We find that twelve of the dwarf candidates are at approximately the radial velocity of the Perseus Cluster, \textit{v} $\approx$ 5366 km s$^{-1}$ \citep{sr}, with the remaining eleven galaxies with radial velocities placing them outside the cluster. All cluster members are within 2.5$\sigma$ of the mean value of the radial velocity obtained by \citet{sr}. Several of the non-cluster members show strong emission lines in their spectra. Interestingly, a number of the background galaxies are at the same redshift ($z$ $\approx$ 0.23) and could be members of a background cluster. The positions and redshifts of these galaxies are listed in Table 2.

The mean radial velocity for the dwarfs is 5750 $\pm$ 43 km s$^{-1}$, slightly higher than the value of $\approx$ 5366 km s$^{-1}$ found by \citet{sr}. \citet{sr} found a velocity dispersion $\sigma$ = 1324 km s$^{-1}$ for galaxies in the Perseus Cluster, and we re-measure this value using just our newly confirmed dwarf galaxy members. We find a velocity dispersion for the dwarfs of $\sigma$ = 1818 km s$^{-1}$, again higher than that found by \citet{sr} for bright galaxies in the Perseus Cluster. The Perseus Cluster however has a complex X-ray morphology \citep{fab81}, and is also not a simple, relaxed system. 

\citet{c3} find a similar result for the dwarf elliptical population of the Virgo Cluster, where the velocity dispersion of the ellipticals is 462 km s$^{-1}$, whilst the velocity dispersion of the dwarf elliptical population is 726 km s$^{-1}$. The fact that the velocity dispersion of dwarf ellipticals is higher than the overall cluster is potentially a sign that the dwarfs may originate from infalling galaxies. Dwarf ellipticals could originate from spiral galaxies that are stripped of mass by high-speed interactions. These dwarfs would retain the velocity characteristics of the original infalling spirals, having velocity dispersions higher than the cluster ellipticals, reflecting the mass distribution of the cluster when the galaxies were accreted. Dwarfs that lie in the central parts of clusters are also susceptible to cannibalism and destruction. Therefore dwarfs would be found preferentially in the outer parts of clusters at high velocity dispersion as these are the most likely dwarfs to survive in the cluster environment. Also, we cannot rule out mass segregation occurring \citep{c3}.

\subsubsection{\textit{HST} ACS imaging}

Deep $HST$ ACS imaging reveals detail not seen in the ground based WIYN imaging (Figures 1 and 2). From such images, galaxies such as background spirals are more easily identified. All the confirmed cluster members are elliptical in shape, with no evidence for internal sub-structure (Fig. 1). It is clear from Fig. 2 that three of the non-members (CGW 19, 27 and 37) are spiral galaxies, although this detail is not seen in the original WIYN imaging presented in \citet{c1}, and as a result were incorrectly classified as potential cluster members.

\begin{figure*}
\includegraphics[width=150mm]{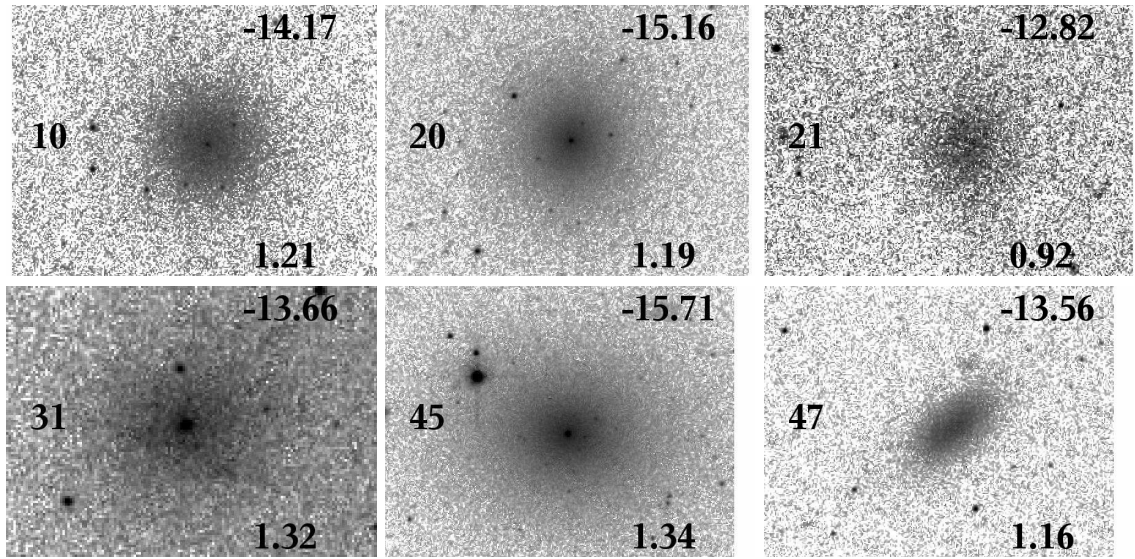}
\caption{Montage of $HST$ ACS images of the confirmed Perseus Cluster dwarfs (galaxies 10, 20, 21, 31, 45, and 47). The number at the upper right of each image is M$_{\rm B}$, the number on the bottom right is the (B$-$R)$_{0}$ colour, and the number on the left is the galaxy number from \citet{c1}.}
\end{figure*}

\begin{figure*}
\includegraphics[width=150mm]{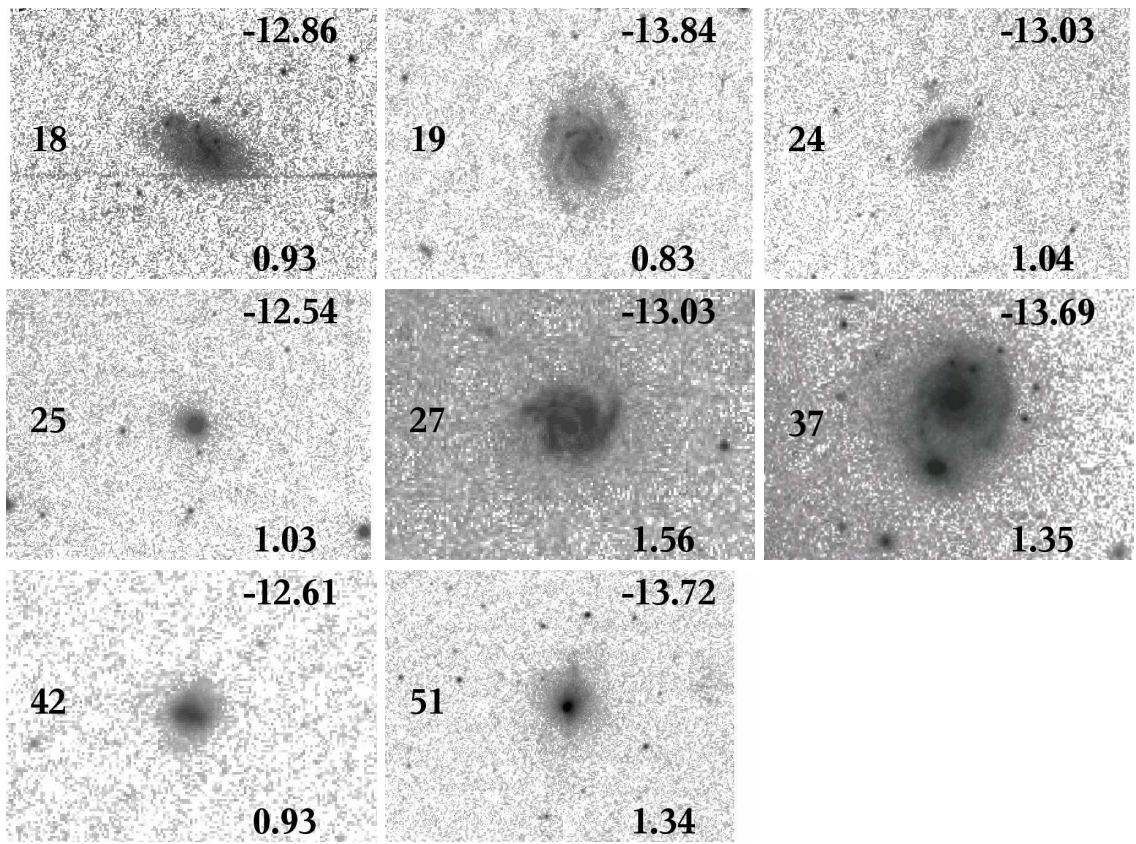}
\caption{Montage of $HST$ ACS images of the non-members (galaxies 18, 19, 24, 25, 27, 37, 42, and 51). The number at the upper right of each image is M$_{\rm B}$ (assuming the galaxy is at the distance of Perseus), the number on the bottom right is the (B$-$R)$_{0}$ colour, and the number on the left is the galaxy number from \citet{c1}.}
\end{figure*}

\subsection{The colour-magnitude relation}

There exists a well-defined colour-magnitude relationship amongst the non-dwarf early-type galaxy population in galaxy clusters, such that brighter galaxies are redder. This correlation is generally regarded as a relationship between a galaxy's mass (traced by magnitude) and metallicity (traced by colour). Using our newly confirmed cluster dwarfs, we investigate whether this relationship extends to the dwarf galaxy population of the Perseus Cluster. Our photometry is taken from \citet{c1}, as well as including some additional photometry on objects not included in \citet{c1}. 

Our colour-magnitude diagram is shown in Figure 3, with the solid line a least-squares fit found by \citet{c1} to the bright cluster ellipticals, and the dashed line an extension of this fit to fainter magnitudes. We carry out a least squares fit to this data using only confirmed members to establish a relation between the colour and magnitude for the confirmed population of Perseus as a whole. This is shown as the dotted line on the colour-magnitude diagram. This fit is slightly steeper than the relationship found by \citet{c2} for the brighter cluster galaxies in Perseus.

The scatter in the colour magnitude relation for the cluster members is small ($\sigma$ = 0.05) down to M$_{\rm B}$ = $-15.5$, but increases to $\sigma$ = 0.11 at magnitudes fainter than this. The degree of colour scatter for the cluster member fainter than M$_{\rm B}$ = $-15.5$ is real, based on a comparison to the photometric errors, to a 2$\sigma$ confidence level. The confirmed faint cluster members that depart from the colour-magnitude relation (CMR), tend to be a bluer than the CMR. These bluer dwarfs are most likely metal poor, as spectroscopic results from previous studies have shown (\citealt{HM}, \citealt{LMF}). The redder colours of two of the dwarf galaxies implies that they may have gone through a process of metal enrichment \citep{c3}. These dwarfs are also likely part of a dwarf galaxy population that have a different origin to the blue dwarfs. \citet{c1} suggest that the most likely origin for these red dwarfs is from infalling spirals. The red cluster dwarfs are most likely not primordial objects, and would have intermediate or young stellar populations.

From the colour magnitude diagram, it appears that galaxies with (B$-$R)$_{0}$ $>$ 1.6 (i.e. the reddest elliptical) are not cluster members. We find no confirmed cluster members with (B$-$R)$_{0}$ $>$ 1.6, although there are non-cluster members with (B$-$R)$_{0}$ less than this, so cluster membership cannot be determined by colour and morphology alone for more distant clusters such as Perseus. Radial velocities obtained via spectroscopy are required as non-cluster members can have colours and structures similar to cluster members, as can be seen on the colour-magnitude diagram in Fig. 3. However, morphology can still be a useful tool in determining cluster membership in more distant clusters from high resolution imaging, as our $HST$ observations have shown.

Broadband (B$-$R)$_0$ colours also reveal information about the nature of the stellar populations in dwarf galaxies, and scatter from the colour-magnitude relationship can be partly explained by the amount of metal enrichment that has occurred. A broad metallicity distribution has been inferred for dwarfs in other galaxy populations such as Fornax and Coma (e.g. \citealt{Po,RS}), where a breakdown in the colour-magnitude relation is seen below M$_{5550}$ $= -17$. Age effects are also likely important in the colours of low-mass dwarf galaxies, with chaotic star formation histories in such galaxies. We investigate this further in $\S$ 3.4 using absorption line indices to derive luminosity weighted ages and metallicities for our dwarfs. 

\begin{figure}
\includegraphics[width=84mm]{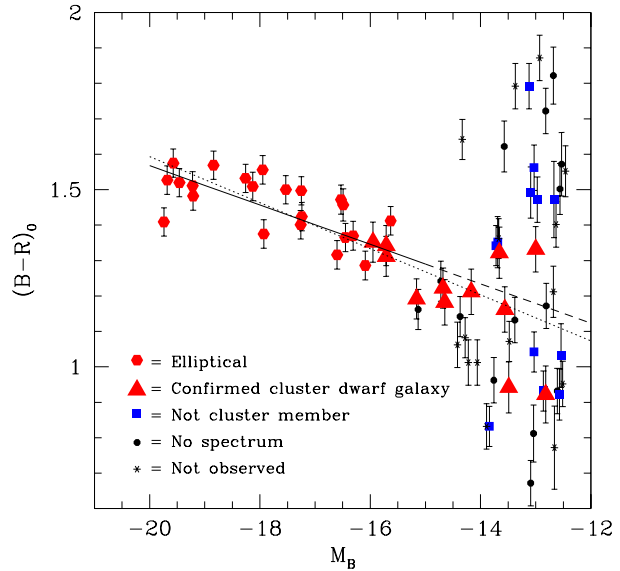}
\caption{Colour magnitude diagram for dwarf galaxies in the Perseus Cluster, taken from \citet{c1}. We include all objects with M$_{\rm B}$ $<$ $-12.5$. All the dwarfs listed in Table 1 are included in this plot. The solid line represents the fit between colour and magnitude for the objects with M$_{\rm B}$ $<$ $-16$, with the dashed line the extension of the fit to fainter magnitudes. The dotted line is the colour-magnitude relation for the cluster ellipticals and dwarf galaxies.}
\end{figure}

\subsection{The central luminosity function}

One of the main reasons for fitting luminosity functions is to determine the relative number of faint systems compared to brighter galaxies. This is typically done by measuring the value of the faint-end slope $\alpha$, with values typically in the range $\approx$ $-1$ and $-2.3$ in clusters (\citealt{c2,mie,bei}).

Previous to this study, values were found for the faint-end slope $\alpha$ $= -1.56$ $\pm$ 0.07 and $-1.44$ $\pm$ 0.04 (\citealt{dpp,c2}) in the central regions of the Perseus Cluster. We refit this luminosity function after removing all galaxies now classified as non-cluster members, and those with (B$-$R)$_{0}$ $>$ 1.6 (the colour of the reddest elliptical) from the luminosity function in \citet{c2}.  Galaxies with (B$-$R)$_{0}$ $>$ 1.6 are not included in the luminosity function as they are most likely not cluster members. 

Figure 4 shows our new luminosity function to a limiting magnitude of M$_{\rm B} -12.5$. We fit this luminosity function with a power law, of the form dN/dL = L$^{\alpha}$. Galaxies are binned according to their magnitude, and a fit was made to the number counts using a least-squares. The resulting fit is shown in Figure 4 as a solid line. We find the faint end slope to be $\alpha$ $= -1.26$ $\pm$ 0.06. This value is flatter than that found previously by \citet{c2}.

\begin{figure}
\includegraphics[width=84mm]{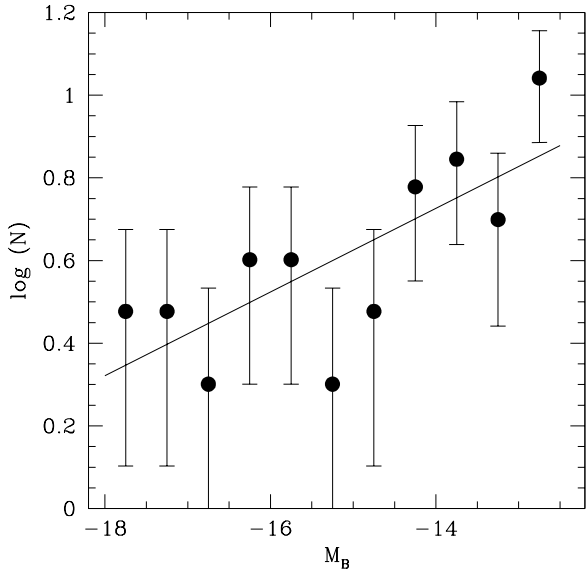}
\caption{The luminosity function for galaxies with M$_{\rm B}$ $>$ $-18$ and (B$-$R)$_{0}$ $<$ 1.6 in the central region of Perseus. The solid line is the least-squares fit to the data.} 
\end{figure}

Importantly, our value of $\alpha$ also matches that of the field, with \citet*{trent} finding an average logarithmic fit of $\alpha$ $= -1.26$ $\pm$ 0.11 for the field luminosity function. This result implies that there is no difference between the relative density of dwarf galaxies to giant galaxies in the field and in the centre of the Perseus Cluster.

It is worth mentioning that the determination of the faint end slope is strongly dependent on the technique used to derive the luminosity function, in particular how the cluster members are determined. The selection process for cluster members as outlined in \citet{c2} along with the spectroscopic technique for selecting cluster members in this study has led to us removing all dwarfs with (B$-$R)$_{0}$ $>$ 1.6. Other methods used to determine cluster membership include selecting dwarfs according to their morphology and surface brightness profiles (e.g. \citealt{trenth02}). Another method is to perform a background subtraction, with the luminosity function determined from a measurement of the excess of galaxies relative to the field \citet{phil98a}. This method can lead to a very steep faint end slope, due to contamination from background galaxies.

\citet{moore} explain the dwarf-density relation as the result of galaxy harassment. Harassment operates more effectively in denser regions, such as cluster cores. Tidal effects become increasingly important in the inner regions of clusters, with smaller galaxies more easily destroyed in cluster centres. This could explain why we find a faint-end slope $\alpha$ equal to that of the field, as the lowest mass galaxies are simply not able to survive in the innermost regions of galaxy clusters.

\subsection{Ages \& metallicities from Lick indices}

Measurements of the ages and metallicities of the stellar populations in our dwarfs are made using the spectra of the newly confirmed members. The strength of the Balmer absorption lines in a stellar population reflect the luminosity-weighted effective temperature of the system, which is dominated by the main-sequence turnoff (e.g. \citealt{sm}). In a single stellar population (SSP) model, the tun-off luminosity and temperature depend on the age of the stellar population.

To measure the equivalent widths if our absorption lines we follow the procedure outlined in \citet{BH} who measured the indices of globular clusters with low-resolution spectra. We measure Lick indices from our spectra to determine ages and metallicities for our dwarfs (\citealt{wo,trea,nelan}), although it was not possible to measure every absorption feature for each galaxy. These indices are calculated with respect to pseudo-continua, defined on either side of the feature bandpass. We can be sure that there is no star formation occurring in our dwarfs based on deep H$\alpha$ narrow-band imaging of these galaxies \citep*{cgw01b}.

The line-strength indices are computed by first shifting the galaxy spectra to their rest-frame wavelengths. We did not flux calibrate our data, but flux calibrated indices differ from non-flux calibrated ones by $<$1$\%$, and this does not contribute significantly to our errors \citep{str}. The spectra were then smoothed using a wavelength-dependent Gaussian kernel to match the Lick/IDS resolution of $\sim$8-10 \AA\ \citep{wo}.

In our analysis, we measured the the Balmer line indices H$\alpha_A$, H$\alpha_F$, H$\beta$, and the metal line indices Mg \textit{b}, Fe $\lambda$5270 and Fe $\lambda$5335 using the passband definitions in \citet{trea} and \citet{nelan}. We tested our method on model spectra provided online by Guy Worthey, and our line measurements matched the expected ones to $<$ 5$\%$. We were unable to obtain Lick standards during our observations, so we are not able to provide an offset between our observations and the Lick System. However, when \citet{bro} measured the Lick indices on six standard stars under an identical LRIS set-up, no significant deviations from the Lick system were found. The offsets found are typically smaller than the index errors that they measured, so were not applied to the indices. Therefore, it was not necessary to apply an offset between our observations and the Lick System. 

The strengths of metal lines can be influenced by the ages of the stellar populations, as well as by their metallicity. This complicates the process of constraining the metallicities of the dwarf galaxies. However, when a Balmer index is plotted against a metal line, it yields a two-dimensional theoretical grid from which the luminosity weighted ages and metallicities can be estimated. Several models exist that predict the Lick indices for SSPs at various metallicities and ages, with the models used throughout this paper being those of \citet*{tho} and \citet{sm}. The \citet{tho} models cover lines in the wavelength range 4000 {\rm \AA} $\lesssim$ $\lambda$ $\lesssim$ 6500 {\rm \AA}, which is not sufficient to cover the H$\alpha$ line. We use the \citet{sm} models to compare with this index.

\subsubsection{Ages and metallicities}

Before deriving the ages and metallicities for our Perseus dwarfs, we estimated the $[\alpha/{\rm Fe}]$ values for our galaxies. The $\alpha$ elements are created rapidly in Type II supernovae, while iron is produced by Type Ia supernovae on longer timescales. By examining the location of our dwarf galaxies on a plot of $<$Fe$>$ versus Mg $b$, we can derive likely values of $[\alpha/{\rm Fe}]$ for our cluster dwarfs (Fig 5). Mg $b$ is an indicator of the $\alpha$ elements, and $<$Fe$>$ =  $\frac{1}{2}$(Fe5270 $+$ Fe5335). 

\begin{figure}
\includegraphics[width=84mm]{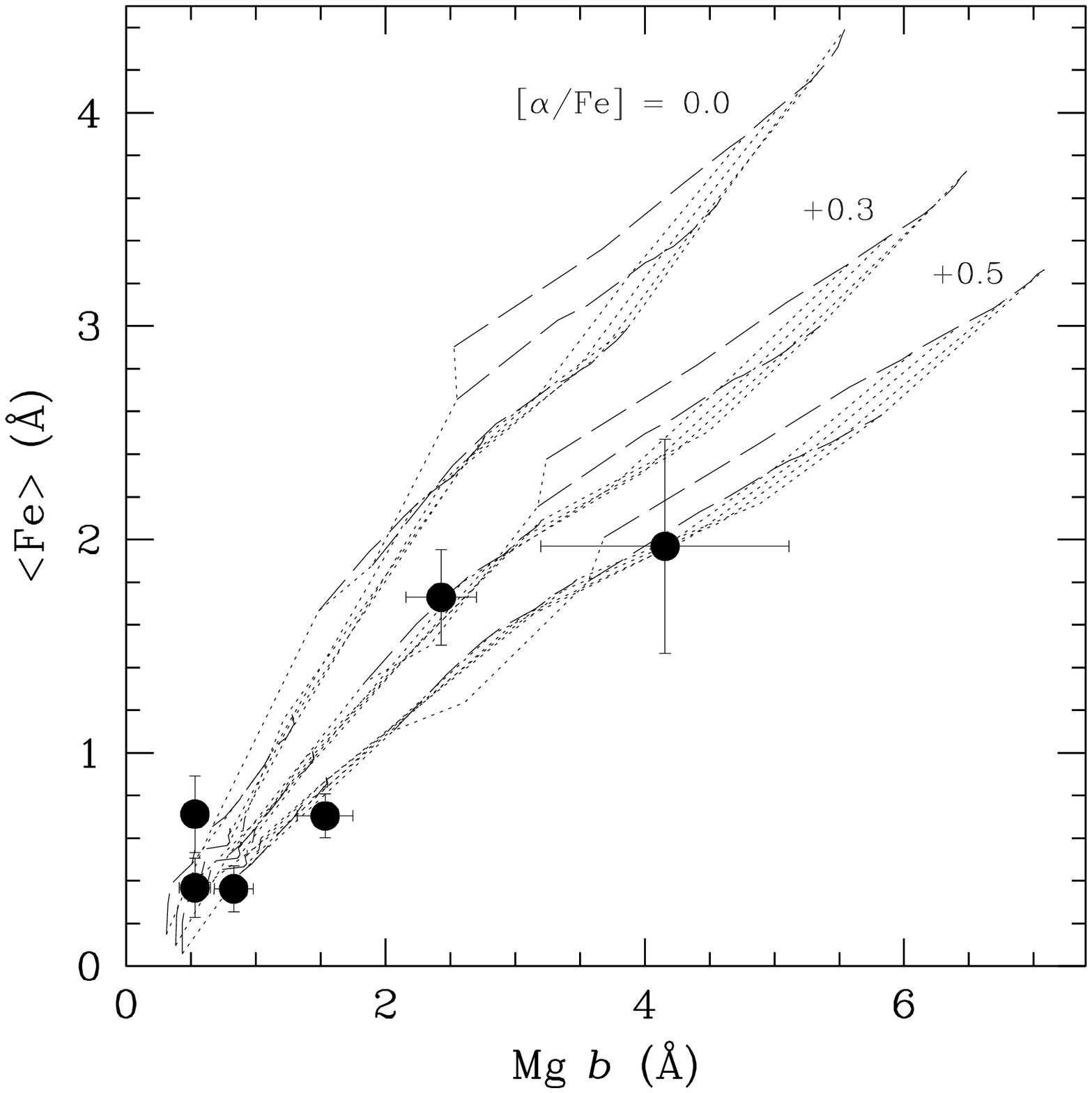}
\caption{Relationship between the $<$Fe$>$ index and the Mg $b$ index for the cluster dwarfs. Model predictions by \citet{tho} for the relationship between $<$Fe$>$ and Mg $b$ are plotted for the abundance ratios $[\alpha/{\rm Fe}]$ = 0.0, +0.3 and +0.5. The dotted lines represent the model predictions for the ages 1, 5, 8, 11 and 15 Gyr, with age increasing from left to right. The dashed lines represent the model metallicities of -2.25, -1.35, -0.33, 0.0 and +0.35, with metallicity increasing from bottom to top.}
\end{figure}

We find that two of our dwarfs are consistent with super-solar abundance ratios ($[\alpha/{\rm Fe}]$ $\approx$ +0.3). Super-solar abundances are consistent with older stellar populations, such as globular clusters and elliptical galaxies. Super-solar $[\alpha/{\rm Fe}]$ indicate that enrichment from Type II SNe has taken place. The \citet{tho} models converge at low metallicities, and our errors are not small enough to reject lower $[\alpha/{\rm Fe}]$ ratios. Therefore, we use a value of $[\alpha/{\rm Fe}]$ = 0.3 when determining the ages and metallicities for our dwarfs. Other studies have found $[\alpha/{\rm Fe}]$ ratios as high as this in a dwarf population (e.g. \citealt{con06}).

To measure the ages of our Perseus dwarfs, we first compare our measured H$\alpha$ indices to the \citet{sm} single stellar population models of H$\alpha_A$ and H$\alpha_F$, as a function of age and metallicity (Fig 6), with metallicities [Fe/H] = $-0.38$, 0.00, $+0.32$, and $+0.56$, although we exclude the [Fe/H] = $+0.32$ model for simplicity in Figure 6. These models assume solar-scaled chemical abundance ratios, whereas we have found our dwarfs to have super-solar abundance ratios. However, this model is thought to be more robust against variation in $[\alpha/{\rm Fe}]$ than the higher order Balmer lines. The H$\alpha$ line has a weak response to metallicity variations, so it should be a good indicator of stellar ages in integrated spectra. Several ages are seen within the dwarfs, with some having ages $>$ 5 Gyr, and the remaining a younger population.

\begin{figure}
\includegraphics[width=84mm]{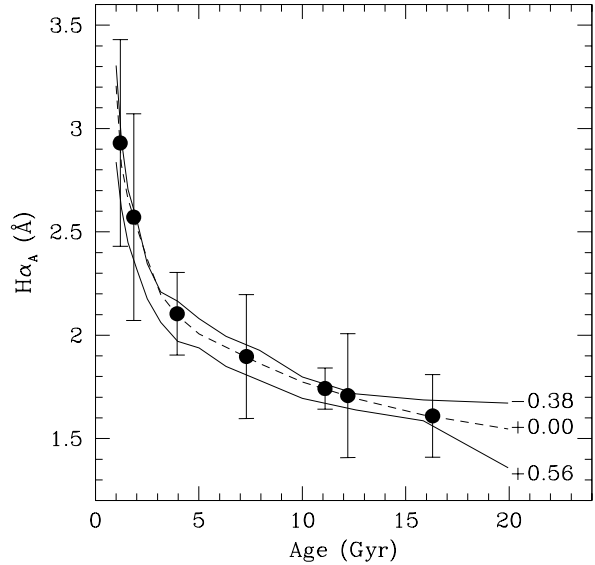}
\caption{Comparison between the H$\alpha_{A}$ indices for our dwarf galaxies and the [Fe/H] = 0.00 model from \citet{sm} (dashed line) of how H$\alpha_{A}$ evolves as a function of time. The models for [Fe/H] = $-0.38$, $+0.56$ are also plotted as solid lines. The points are the measured values of H$\alpha_{a}$ that we obtained from our dwarf spectra plotted at the ages at which these values are expected in the [Fe/H] = 0.00 model}
\end{figure}

By combining a metal index with a Balmer index, we can place constraints on the luminosity-weighted ages and metallicities of the stellar populations of the newly confirmed Perseus dwarfs. We use the Mg $b$ index for this purpose, as well as the $<$Fe$>$ index. 

Figure 7 shows that at least one of the Perseus dwarfs is located in the low-metallicity, and old age part of the H$\beta$-Mg $b$ diagram, with ages $\sim$ 8 $\pm$ 3 Gyr and metallicities $[{\rm Fe/H}]$ $<$ -0.33 at a 3$\sigma$ confidence level. There is also a younger, more metal rich dwarf with age $\sim$ 3 $\pm$ 2 Gyr and $[{\rm Fe/H}]$ $\sim$ 0.35. Comparing the H$\beta$ indices to the $<$Fe$>$ indices produces similar results, with one dwarf galaxy having an age of 11 $\pm$ 3 Gyr, and with a metallicity less than [Fe/H] $= -0.33$. Within the error bars, another dwarf is potentially older than 5 Gyr and again has $[{\rm Fe/H}]$ $<$ $-1.35$. One dwarf is clearly much younger ($<$ 5 $\pm$ 3 Gyr) and more metal rich ($[{\rm Fe/H}]$ $>$ $-0.33$. The remaining two galaxies lie below the grid so it is not possible to make estimates for their ages and metallicities, but they are most likely older ($>$ 8 Gyr) and metal poor systems. The differences in age and metallicity measured throughout this paper are real to 2-3$\sigma$.

We also compare the H$\beta$ and the $[\alpha/{\rm Fe}]$-insensitive index [MgFe]' = $[({\rm Mg} $b$)(0.72 \times {\rm Fe}5270+0.28 \times {\rm Fe}5335)]^{0.5}$ \citep{tho}. This index servers as the best tracer of the overall metallicity, H$\beta$ is also less sensitive than the other Balmer line indices to metallicity. Again, we find the same results with two old, metal poor dwarfs ($[{\rm Fe/H}]$ $<$ $-1.35$, age $>$ 8 Gyr), and one metal rich, young dwarf ($[{\rm Fe/H}]$ $>$ $-0.33$, age $<$ 5 Gyr).

We find several H$\beta$ indices that are too low to be fitted by the \citet{tho} models. This has been previously observed in the spectra of dwarf galaxies by \citet{Po}, who suggest that current models overestimate the turnoff temperatures in the models. Such galaxies are interpreted as old stellar systems, with ages $>$ 8 Gyr for the purposes of this paper.

\begin{table*}
\begin{minipage}{168mm}
\caption{Line-strength indices}
\begin{tabular}{lcccccc}
\hline
Galaxy & H$\beta$ & Mg \textit{b} & $<$Fe$>$ & [MgFe]' & H$\alpha_A$ & H$\alpha_F$ \\
\hline
CGW 10 & 2.7 $\pm$ 0.5 & \ldots & \ldots & \ldots & 3.0 $\pm$ 0.5 & 1.9 $\pm$ 0.3\\
CGW 16 & 3.8 $\pm$ 0.7 & 0.8 $\pm$ 0.2 & 0.36 $\pm$ 0.1  & 0.38 $\pm$ 0.2 & \ldots & \ldots \\
CGW 20 & 2.5 $\pm$ 0.4 & 1.5 $\pm$ 0.2 & 0.70 $\pm$ 0.1 & 0.96 $\pm$ 0.3 & 1.6 $\pm$ 0.2 & 1.3 $\pm$ 0.1 \\
CGW 21 & \ldots & \ldots & \ldots & \ldots & \ldots & \ldots \\
CGW 29 & \ldots & 2.5 $\pm$ 0.7 & \ldots & \ldots & 1.0 $\pm$ 0.3 & 1.2 $\pm$ 0.3 \\
CGW 31 & \ldots & 0.5 $\pm$ 0.1 & 0.71 $\pm$ 0.2 & 0.56 $\pm$ 0.2 & 1.7 $\pm$ 0.3 & 1.2 $\pm$ 0.3 \\
CGW 38 & 1.7 $\pm$ 0.2 & 0.5 $\pm$ 0.1 & 0.36 $\pm$ 0.1 & 0.69 $\pm$ 0.2 & 1.7 $\pm$ 0.1 & 1.8 $\pm$ 0.1 \\
CGW 39 & 1.4 $\pm$ 0.2 & 2.4 $\pm$ 0.3 & 1.7 $\pm$ 0.2 & 2.3 $\pm$ 0.4 & 1.1 $\pm$ 0.1 & 2.2 $\pm$ 0.2 \\
CGW 45 & 1.1 $\pm$ 0.2 & 2.5 $\pm$ 0.3 & \ldots & \ldots & 2.1 $\pm$ 0.2 & 1.8 $\pm$ 0.2 \\
CGW 47 & 2.7 $\pm$ 0.5 & \ldots & \ldots & \ldots & 3.0 $\pm$ 0.4 & 2.1 $\pm$ 0.3 \\
CGW 49 & \ldots & \ldots & \ldots & \ldots & 1.9 $\pm$ 0.3 & 1.6 $\pm$ 0.2 \\
54 & 2.35 $\pm$ 0.6 & 4.2 $\pm$ 1.0 & 1.9 $\pm$ 0.5 & 2.8 $\pm$ 1.2 & 2.6 $\pm$ 0.5 & 2.2 $\pm$ 0.4 \\ 
\hline
\end{tabular}

\medskip
Line-strength indices as determined by the definitions in \citet{trea} and \citet{nelan}. The combined iron index is defined as $<$Fe$>$ = $\frac{1}{2}$(Fe5270+Fe5335) and the $[\alpha/{\rm Fe}]$-insensitive index [MgFe]' is defined as [MgFe]' = $[({\rm Mg} $b$)(0.72 \times {\rm Fe}5270+0.28 \times {\rm Fe}5335)]^{0.5}$ \citep{tho}. Not every index was measured for every galaxy.
\end{minipage}
\end{table*}

\begin{figure*}
\includegraphics[width=171mm]{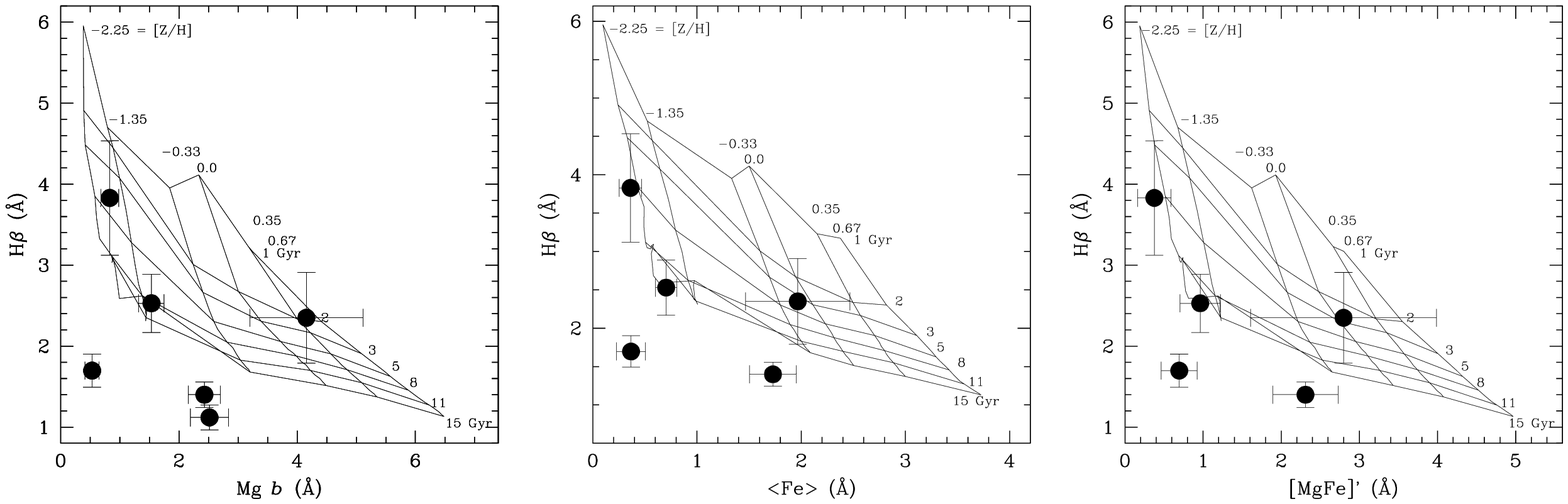}
\caption{Plot of H$\beta$ versus Mg $b$, $<$Fe$>$ and [MgFe]' \textit{(left to right respectively)} for the Perseus dwarfs superimposed on \citet{tho} model isochrones and isometallicity lines for $[\alpha/{\rm Fe}]$ = 0.3. The different ages and metallicities are labelled.}
\end{figure*}

\subsection{Comparison of colour with age and metallicity}

Dwarf ellipticals without star formation that are blue in colour are expected to be metal poor, whereas those with redder colours are expected to be metal rich. Unfortunately, we are unable to derive ages and metallicities for the two redder dwarfs (galaxies CGW 31 and CGW 49), as not all the spectral features could be measured reliably.

One galaxy, 54 (Table 1), is younger and more metal rich than the other dwarfs for which we obtained a good set of indices. It has an age $<$ 5 Gyr, and a metallicity [Fe/H] $>$ -0.33. The (B$-$R)$_{0}$ colour of this particular galaxy is 1.22, and it lies on the fit between colour and magnitude for the cluster ellipticals and dwarf galaxies. The other galaxies for which ages and metallicities are determined are CGW 16, and CGW 20. These galaxies are metal poor and have (B$-$R)$_{0}$ colours of 1.18, 1.19 and 1.16 respectively. These galaxies again lie close to the colour-magnitude relation for the cluster ellipticals and dwarf galaxies. Therefore it appears that both young and old dwarf galaxies in Perseus lie on the same colour-magnitude relation. This is confirmed with ages obtained using the H$\alpha_A$ index, when again both young and old dwarf galaxies in Perseus lie on the same colour-magnitude relation. 

\begin{figure*}
\includegraphics[width=171mm]{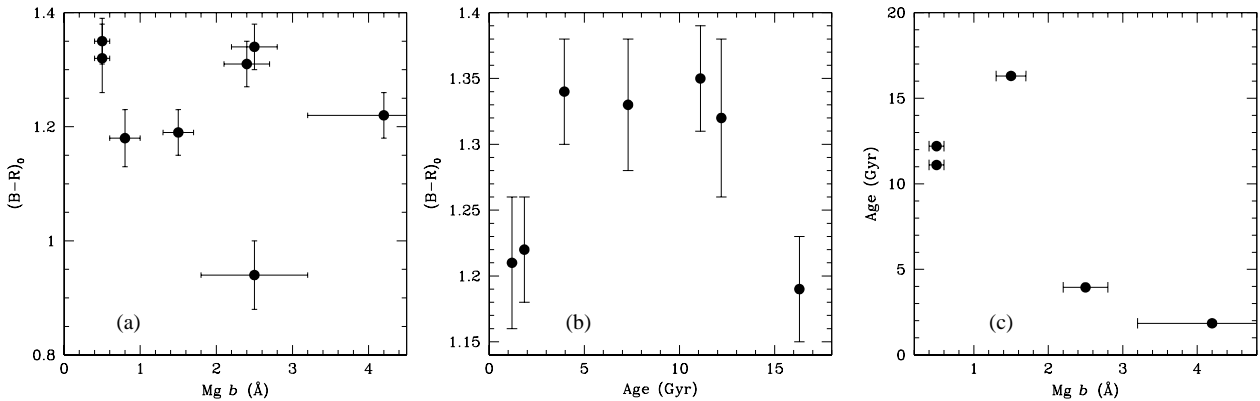}
\caption{$(a)$ Plot of (B$-$R)$_0$ colour against the Mg $b$ index to show how metallicity varies with colour. $(b)$ (B$-$R)$_0$ colour plotted against ages determined using the H$\alpha_{A}$ index. $(c)$ Plot of age against the Mg $b$ index to show how age varies with metallicity.} 
\end{figure*}

We find no correlation between either metallicity or age, determined using the H$\alpha_{A}$ index, with (B$-$R)$_{0}$ colour as shown in Figures 8$a$ and 8$b$. However, we do find a clear anti-correlation between age and metallicity (Fig. 8$c$). The older the dwarf elliptical, the more metal poor it is. This has been found in previous studies such as that of \citet{Po}, who find an anti-correlation between age and metallicity for galaxies in the Coma Cluster in any luminosity bin. 

\subsection{Irregular galaxy}

\begin{figure}
\includegraphics[width=84mm]{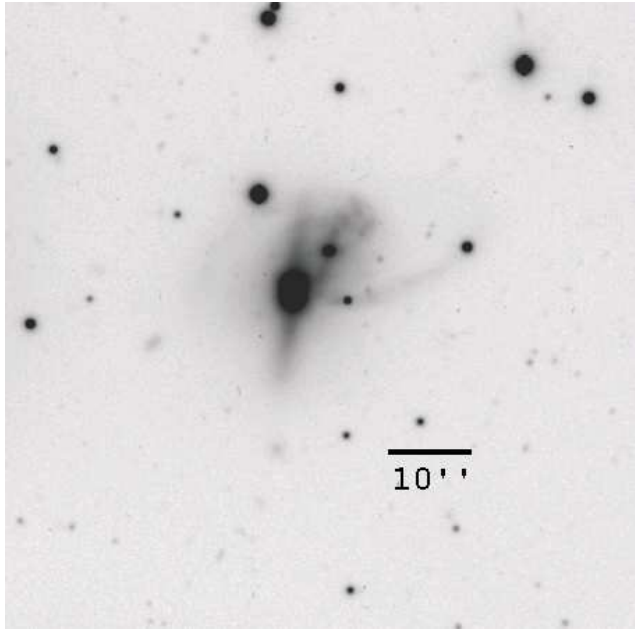}
\caption{WIYN Harris $R$ band image of the irregular galaxy at $\alpha$ = 03 19 05.1, $\delta$ = $+$41 28 12.4. The solid black line in this figure is 10'' in length.}
\end{figure}

We also obtained a spectrum of an irregular galaxy at $\alpha$ = 03 19 05.1 and $\delta$ = $+$41 28 12.4. From the absorption lines, this galaxy was found to be at a radial velocity of 4032 $\pm$ 39 km s$^{-1}$, confirming it as a member of the Perseus Cluster. A second object was present in the slit for this galaxy, which had Balmer lines in emission in its spectrum. Based on the positions of these emission lines, the additional object was found to be at $z$ $\approx$ 0.055, giving the object a radial velocity of $\approx$ 16500 km s$^{-1}$. This would make it either a background galaxy, or an interacting galaxy with a velocity difference of $\approx$ 12450 km s$^{-1}$ from the main galaxy.

From the image of this galaxy, a tidal tail is clearly visible to the left hand side of the galaxy as seen in Figure 9. This irregular galaxy is therefore most likely a merged system. Assuming the Perseus Cluster is at a distance of 77 Mpc, this tidal tail has a length of $\approx$ 7.5 kpc. 

\section{Interpretation and Discussion}

We find that the radial velocity dispersion for dwarf galaxies, $\sigma =$ 1818 km s$^{-1}$, is higher than the value of 1324 km s$^{-1}$ found by \citet{sr} for the giant galaxies in Perseus. These results agree with results found by \citet{c3} for dwarf galaxies in the Virgo Cluster. An infall origin for dwarfs can explain this difference in velocity dispersion between the brighter cluster galaxies and the dwarfs. Infalling spirals can become morphologically transformed into dwarf ellipticals by high speed interactions with other cluster galaxies, and by the gravitational potential of the cluster itself (harassment). Another possibility is that infalling dark matter halos enter the cluster at later times, with the gas in the halos being converted into stars due to tidal forces during the cluster assembly. The velocity characteristics of low-mass, infalling galaxies should change little over several gigayears, thus the velocity dispersion would represent the mass profile of the cluster at the time of accretion. 

We also find in Perseus an increase in the scatter of the colour-magnitude relation for magnitudes fainter that M$_{\rm B}$ $= -15.5$, which can be interpreted as a diversity in the ages and metallicities of the low mass systems. A range in the ages and metallicities of low-mass systems suggests that dwarf galaxies are not a homogeneous population, as redder, more metal rich cluster dwarfs must have been formed more recently than the blue population. 

\citet{DeL} suggest that a large fraction of faint red galaxies in current clusters moved onto the colour-magnitude relation relatively recently, with their star formation activity coming to an end at around z $\approx$ 0.8. Clusters at redshift z $\approx$ 1 have a deficit of red-sequence galaxies, with the dwarf cluster population becoming progressively fewer in number as we move out to increasing redshifts. The rate of infalling spiral galaxies is expected to peak at $z$ $\approx$ 0.8 in a CDM cosmology with $\Omega_{M}$ = 1 and $H_{0}$ = 50 km s$^{-1}$ \citep{Kauf}, the same redshift at which \citet{DeL} suggest a large fraction of faint red galaxies in current clusters moved onto the colour-magnitude relation. If dEs are indeed formed from infalling spirals, it seems likely that they originate from this epoch.

We find that the faint-end luminosity function is less steep than what was found in previous studies of the Perseus Cluster, and in fact matches the value of $\alpha$ $= -1.26$ found by \citet{trent} for the field. We remove all galaxies incorrectly classified as dwarfs by \citet{c1} from the luminosity function, to refit this faint end slope. 

The shape of the luminosity function at faint magnitudes is dependent on many processes, one of which is the efficiency at which gas is converted into stars (e.g. \citet{ds}. Gas loss from supernova driven winds would reduce the star formation efficiency. Star formation and/or nuclear activity in dwarf galaxies could also photoionize the intergalactic medium, regulating the galaxy formation process \citep{est92}. Another effect on the faint-end slope is that dark halos in denser, evolved clusters are more efficient at collecting gas used to form stars than those in unevolved clusters \citep{trenth02}. Yet another process changing the shape of the faint end slop is dynamical stripping of higher mass galaxies that would increase the number of low-mass systems in a cluster \citep{con02}. Also, dwarfs in the inner regions of dense clusters are less able to survive due to harassment \citep{moore}, causing the faint-end slope to flatten for such environments. 

Whilst our value of $\alpha$ $= -1.26$ $\pm$ 0.06 for dwarfs in the Perseus Cluster core is comparable with that of the field, it is also similar to the value of $\alpha = -1.23 \pm 0.13$ \citep{Pra} in the core region of Abell 2218. In Abell 2218, the slope of the luminosity function varies with cluster environment, going from $\alpha$ $= -1.23$ $\pm$ 0.13 in the central core of the cluster to $\alpha$ $= -1.49$ $\pm$ 0.06 outside of the core. \citet{Pra} infer that the core of Abell 2218 is ``dwarf depleted", with the dwarf-to-giant ratio decreasing monotonically with increasing cluster density. Galaxy clusters with less prominent dwarf populations are in fact those with the highest projected galaxy densities \citep{phil}.

The reason for this depletion in the core regions of clusters can perhaps be explained by galaxy harassment. Tidal effects become more important in the centre of clusters, with smaller spheroidal galaxies being easily destroyed \citep{con02}. The lowest surface brightness objects will disintegrate in the inner regions, decreasing the dwarf-to-giant ratio, whilst in outer regions such galaxies would be able to survive. 

More recently, \citet{hdp} find no evidence for a faint-end upturn in the luminosity function  in clusters at $z$ = 0.3. There will be less contamination from background galaxies in clusters at these redshifts, and therefore such galaxies would not artificially steepen the faint-end slope to the degree seen in local clusters. \citet{hdp} argue that the faint-end upturn in the LF in nearby clusters is of recent origin, although our results could contradict this theory as we find no evidence for a faint-end slope upturn.

A hierarchical model of galaxy formation predicts that within a given environment, low-mass galaxies would be the first to form. However, it appears that many low mass galaxies finish forming after the giants, through ``downsizing". For example, results from \citet{DeL} suggest that a large fraction of dwarf galaxies in current clusters moved onto the red sequence relatively recently, with a cessation in star formation activity at \textit{z} $\ll$ 0.8. This is also seen for field galaxies \citep{bun}, suggesting downsizing is a possible origin for dwarf galaxies in both environments. We find both old, metal-poor dwarfs and young, metal rich dwarfs in Perseus. This is consistent with results by \citet{Po} for faint galaxies in Coma, where dwarf galaxies occupy a large region on the SSP model grid, covering a range of metallicity and ages, with most lying in the region [Fe/H] $<$ $-0.25$ and ages $<$ 1 Gyr.

These results imply that downsizing is a real effect, with the younger dwarfs ending star formation at some time after the giant cluster members. Whilst some dwarfs are indeed primordial, not all cluster dwarfs are old. The formation of these young dwarfs may involve the removal of mass from infalling higher-mass spirals (e.g. \citealt{c1}).

\section{Conclusions}

We have analysed Keck/LRIS spectra for twenty-three dwarf galaxy candidates in the Perseus Cluster, from which we have confirmed cluster membership for twelve systems based on radial velocities measured from absorption lines.

We extend the confirmed member colour-magnitude relation for Perseus down to M$_{\rm B}$ $= -12.5$, finding that the slope of the colour-magnitude relation becomes bluer when the low-surface brightness dwarf galaxies are included. The fainter dwarfs also scatter more from the colour-magnitude relation, following the trend observed by \citet{RS} for low-mass galaxies in Fornax and Coma. This scatter can be interpreted as a spread in the metallicity distributions of dwarf galaxies, which has been inferred for dwarf galaxies in other clusters by \citet{RSMP} and \citet{Po}.

After removing non-members from the $B$-band luminosity function for the Perseus Cluster we find a faint-end slope $\alpha$ $= -1.26$ $\pm$ 0.06, similar to the field. Previous studies of other galaxy clusters have found that the dwarf-to-giant ratio is a function of local projected galaxy density. Extending this study to the outer regions of Perseus would enable us to see how the luminosity function changes with the local galaxy density. 

Colours, morphologies, and central surface brightnesses are not sufficient criteria to confirm cluster membership, as this work has shown. Cluster membership cannot be confirmed without spectroscopy, so the faint-end luminosity functions calculated for galaxy clusters where membership has not been confirmed spectroscopically are most likely artificially steepened by non-members resulting in higher dwarf-to-giant ratios.

By comparing the observed dwarf spectral absorption indices with population synthesis models of \citet{tho}, we derive luminosity-weighted ages and metallicities for the dwarf galaxies. A range of ages is observed, ranging from older than 8 Gyr to younger than 5 Gyr. The metallicity distribution of the faint cluster members is also not that of a simple homogeneous population, with the younger galaxies typically having higher metallicities.

More observations of dwarf galaxies in rich, nearby cluster environments are required in order to help improve our understanding of the formation scenarios for cluster dwarfs. Further spectroscopic work with a larger sample would also allow better constraints on the ages and metallicities of cluster dwarf galaxies and would help with modelling the formation of such galaxies.

\section*{Acknowledgments}

This research has made use of the NASA/IPAC Extragalactic Database (NED) which is operated by the Jet Propulsion Laboratory, California Institute of Technology, under contract with the National Aeronautics and Space Administration. We thank the staff of Keck, WIYN and NOAO for help in obtaining the spectroscopy and imaging presented here. We also thank Jay Gallagher for help with the WIYN optical imaging, and Dolf Michielsen for useful advice. S. J. P. acknowledges the support of a PPARC studentship. We also wish to acknowledge the highly significant cultural role and reverence that the indigenous Hawaiian community hold for the summit of Mauna Kea. It is a privilege to be able to conduct observations from this mountain.

\end{document}